\documentclass[man,noextraspace]{apa6} 
\usepackage{IEEEtrantools}
\usepackage[english]{babel}
\usepackage[utf8x]{inputenc}
  \usepackage{setspace}
\usepackage{amsmath,amsthm,amssymb}
\usepackage{graphicx}
\usepackage{bm}
\usepackage[colorinlistoftodos]{todonotes}
\usepackage[authoryear]{natbib} 
\bibpunct{(}{)}{;}{a}{,}{,} 
\usepackage[flushleft]{threeparttable}
\usepackage{booktabs,caption}
\usepackage{colortbl}
\usepackage{graphicx}
\usepackage{pdflscape}
\usepackage{caption}
\usepackage{subcaption}
\usepackage{enumitem}
\usepackage{graphicx}
\usepackage{bm}
\usepackage{textcomp}
\usepackage{textgreek}
\usepackage{caption}
\usepackage{multicol}
\usepackage{IEEEtrantools}
\usepackage{booktabs}
\usepackage{authblk}
\usepackage{outline}
\usepackage{mathrsfs}
\usepackage{mathdesign}
\usepackage{euscript}
\usepackage{amsbsy}
\usepackage{mathtools}
\usepackage{multirow}
\usepackage{rotating}
\usepackage[ruled,vlined]{algorithm2e}
\SetKwRepeat{Do}{do}{while}%
\usepackage{array}

\usepackage{listings}
\DeclareMathOperator*{\argmin}{argmin}
\DeclareMathAlphabet{\mathscrbf}{OMS}{mdugm}{b}{n}
\newtheorem*{@remark}{\bf Remark}

\title{Structured Estimation of Heterogeneous Time Series}

\shorttitle{multivar II}
\author[1]{Zachary F. Fisher}
\author[2]{Younghoon Kim}
\author[2]{Vladas Pipiras}
\author[1]{Christopher Crawford}
\author[1]{Daniel J. Petrie}
\author[1]{Michael D. Hunter}
\author[3]{Charles F. Geier}

\affil[1]{The Pennsylvania State University}
\affil[2]{University of North Carolina at Chapel Hill}
\affil[3]{The University of Georgia}

\abstract{How best to model structurally heterogeneous processes is a foundational question in the social, health and behavioral sciences. Recently, \citet{fisher2022} introduced the multi-VAR approach for simultaneously estimating multiple-subject multivariate time series characterized by common and individualizing features using penalized estimation. This approach differs from many popular modeling approaches for multiple-subject time series in that qualitative and quantitative differences in a large number of individual dynamics are well-accommodated. The current work extends the multi-VAR framework to include new adaptive weighting schemes that greatly improve estimation performance. In a small set of simulation studies we compare adaptive multi-VAR with these new penalty weights to common alternative estimators in terms of path recovery and bias. Furthermore, we provide toy examples and code demonstrating the utility of multi-VAR under different heterogeneity regimes using the \texttt{multivar} package for R \citep{multivar}. }

\begin{document}
\makeatletter
\@ifundefined{@affil}{\def\@affil{~}{}}
\makeatother
\maketitle

Over the last 50 years Peter Molenaar's work has reimagined the foundation of scientific psychology with an eye toward the persistent heterogeneity of human behavior. Peter's work has shown that even under near-constant genetic and environmental conditions heterogeneity in behavior persists \citep{molenaar1993}. Furthermore, this variability is a fundamental feature of many psychological processes and critical for accurately characterizing and intervening on human behavior \citep{molenaar2004}. Although Peter's work helped to disentangle and distill the idiographic and nomothetic perspectives in psychology, it also firmly located the individual as being critical for a generalizable science of behavior \citep{gates2012}. In this paper we continue the tradition of reconciling nomothetic and idiographic orientations and present a novel approach for modeling time-dependent processes where individuals may differ qualitatively and quantitatively in terms of their underlying dynamics. 

\section{Introduction}

It is widely recognized that many behavioral and biological processes are best understood from a complex systems perspective where numerous factors shape outcomes across the lifespan. These interdependent factors bridge multiple levels of analysis from the biological to the behavioral - producing complex, heterogenous interactions that unfold across time. For many years the data required to support this type of analysis were largely unavailable. However, technological advances have made the collection of intensive time-series data easier than ever. This includes sensor-based physiological measurements, neural activity, health and movement data and measures of emotional states, to name just a few. \\
These changes in our data landscape have coincided with an increasing appreciation of heterogeneity as a ubiquitous and defining feature of human behavior. From treatment effects \citep{bryan2021}, to affective science \citep{foster2021} and psychiatric diagnoses \citep{nunes2020}, heterogeneity is increasingly seen as a critical ingredient for understanding causal mechanisms and developing generalizable treatment protocols. Broadly understood, heterogeneity can refer to both between-person and within-person heterogeneity, or the idea that biological and psychological structures underlying an individual’s functioning may differ, both between individuals, and within the same individual across time, respectively. In this work we are primarily concerned with how between-person heterogeneity in dynamic processes is accounted for at the model level, and the remainder of our discussion reflects this focus.\\

In the remainder of the introduction we briefly introduce some existing approaches for handling between-person heterogeneity in time series models. These comparisons are not meant to be exhaustive, only to highlight in broad strokes how heterogeneity is handled by popular time series models. We also discuss Group Iterative Multiple Model Estimation \citep[GIMME;][]{gates2012}, the inspiration for multi-VAR \citet{fisher2022}, and provide some insight on how the approaches differ. Finally, we outline the novel contributions of our current work in extending the multi-VAR framework in a number of important directions.\\

\subsection{Existing Literature}
 
A number of approaches are capable of accounting for heterogeneity in multivariate time series arising from multiple individuals. Here we briefly highlight a few of these approaches. In doing so we make a distinction between two types of heterogeneity at the time series model level: quantitative heterogeneity and qualitative heterogeneity. We say models allow for quantitative heterogeneity when parameters are allowed to vary in magnitude across individuals in the sample (typically according to a distribution). On the other hand, we say models allow for qualitative heterogeneity when the model structure itself is allowed to vary across individuals. For example, individuals might differ in the lag order of their model, the direction and sign of directed relations, or the overall pattern of zero and nonzero parameters. Finally, models may also accommodate both quantitative and qualitative heterogeneity simultaneously.\\

The most popular approaches for modeling multiple-subject time series in psychological research are designed to accommodate quantitative heterogeneity  \citep[e.g. multilevel time series;][]{bringmann2013a, epskamp2018a,lafit2021,li2022}. In these approaches parameters in the model are typically treated as being sampled from a distribution, where fixed and random effects are recovered, and subsequently used to fashion individual-level models from between-person variation in parameter estimates. When all individuals in the sample possess homogenous dynamics, empirical Bayes estimates of the dynamic parameters are generally accurate and reliable \citep{liu2017,liu2018}. Less work has focused on how these estimates perform in situations where individuals meaningfully vary in their dynamics.\\

Another group of approaches account for heterogeneity at the subgroup or cluster level \citep{bulteel2016,takano2021,park2022b,park2022a}. These approaches typically allow for groups of individuals to differ qualitatively in terms of their underlying time series model structure. Approaches vary on whether parameters governing individual-level models within the same cluster can vary qualitatively. Furthermore, any model that implies a Gaussian distribution can be used to define the components of a mixture model. Recently, \citet{hunter2022} extended this idea to state space models where it becomes possible to have state space mixtures with qualitatively different parameterizations across groups, for example.\\

Another approach, designed to allow for both quantitative and qualitative heterogeneity across individual time series models, is GIMME \citep{gates2012}. The GIMME approach is built on the Structural-Vector Autoregressive (S-VAR) model and uses a stepwise model search algorithm based on modification indices \citep[MI;][]{sorbom1989} in conjunction with test statistics developed in the structural equation modeling (SEM) framework. GIMME is designed to recover both group (generalizable) and individual-level (idiosyncratic) models from multivariate time series data and is available in a well-developed R package \citep{gimme}. Recently, the GIMME algorithm has been extended to accommodate the standard and hybrid VAR \citep{luo2022} models, in addition to the original S-VAR. Although not relevant to the current work, there are additional capabilities of GIMME worth mentioning, including the ability to model direct and modulatory effects of tasks \citep{duffy2021}, exogenous variables \citep{arizmendi2021}, data-driven \citep{gates2017unsupervised} and confirmatory \citep{henry2019} subgroups, multiple solutions \citep{beltz2016}  \footnote{In the original GIMME algorithm each individual begins with a null model (all coefficient values are set to zero).  Modification indices are then used to determine which parameter, if freed, would most improve model fit. Multiple solutions refers to the instance where multiple parameters would lead to numerically equivalent improvements in model fit. To address this situation \citep{beltz2016} developed an algorithm, termed multiple solutions GIMME, that accounts for these equivalent solutions in terms of final model selection.}, and latent variables \citep{gates2020latent}.\\

The multi-VAR approach originally described in \citet{fisher2022}, and extended in the current work, is similar in spirit to GIMME in that both produce group and individual-level models from qualitatively and quantitatively heterogeneous time series data. However, beyond this common objective, the approaches share little in common in their operatlization. For example, GIMME induces sparsity in individual-level models via forward-selection, a stepwise procedure where paths are freed according to modification indices until fit indices suggest a well-fitting model has been achieved.\citet{hastie2015} describe a sparse model as having only a small number of nonzero parameters or weights. In the case of GIMME and multi-VAR we are typically concerned with inducing sparsity in the dynamic part of the model. Furthermore, in GIMME individual datasets are analyzed separately and sequentially, and common paths are determined by user-specified thresholds. The multi-VAR approach induces sparsity in individual-level models via structured penalization of common and unique dynamics, where all individual-level models are estimated simultaneously. What constitutes a common and unique path in the multi-VAR framework is determined using competing penalties in the objective function and cross-validation.  Although additional differences exist, a detailed comparison is beyond the scope of this work. Readers interested in a more technical comparison should see \citet[][p. 769-771]{lane2017} for a step-by-step description of the GIMME algorithm and \citet[][p. 6-11]{fisher2022} for the multi-VAR algorithm, including pseudocode.\\

Finally, it is also possible to accommodate both quantitative and qualitative heterogeneity across individual time series model by simply fitting individual-level models separately. In this way each individual-level model is fit without pooling or sharing any information across individuals. In fact, individual-level models represent an important benchmark for any proposed joint-modeling framework. If joint modeling approaches, such as multi-VAR or GIMME, do not show an advantage over individual-level approaches, then sharing of information across individuals provides no additional benefit and one should prefer the most parsimonious modeling approach. For these reasons we also include popular individual-level modeling approaches as comparisons when evaluating the performance of the proposed estimators.

\subsection{Current Work}

\citet{fisher2022} introduced the \emph{multi-VAR} framework, a method for estimating multiple-subject multivariate time series models using structured regularization. The original proposal contained descriptions of a \emph{standard multi-VAR} implementation based on the Least Absolute Shrinkage and Selection Operator \citep[LASSO; ][]{tibshirani1996} and an \emph{adaptive multi-VAR} based on the adaptive LASSO \citep{zou2006}. In the remainder of the paper we use the term multi-VAR to refer to the general modeling framework proposed in \citet{fisher2022}, and when relevant for a given context we specifically refer to the \emph{standard} or \emph{adaptive} variants.\\

In this paper we extend the adaptive multi-VAR framework described by \citet{fisher2022} in a number of meaningful ways.  First, we propose new estimation methods for the adaptive multi-VAR. Specifically, we introduce new methods for computing adaptive weights and a novel cross-validation procedure for selecting the model hyperparameters. Second, we evaluate the utility of these newly proposed weighting schemes and cross-validation procedures in a small simulation study. In this simulation study we evaluate the performance of the proposed approach across various levels of heterogeneity, comparing it to alternative methods, and illustrating its application using an fMRI study example assessing habitual control. Embedded in the paper are illustrations demonstrating how the multi-VAR approach handles commonly encountered modeling situations, with companion code from the \texttt{multivar} package \citep{multivar}.

\section{Vector Autoregressive Model}
The VAR model is a natural starting point for analyzing multivariate time-dependent data. In this section we describe the VAR model as it applies to data collected from a single individual. Estimation of the VAR model is discussed, as well as methods developed to address its inherent overparameterization. Following this discussion of the single-subject VAR we introduce the multi-VAR approach and discuss its advantages when data is available from multiple subjects.\\
\subsection{Single-Subject Vector Autoregressive Model}
VAR models are a natural fit for many psychological applications as they allow for the inclusion of many potentially relevant variables, provide a concise interpretation of lead-lag relations and can be visualized easily using path or network diagrams. Here we consider a multivariate time series for a single individual, $\{\mathbf{X}_{t}\}_{t \in \mathbb{Z}} = \{(X_{j,t})_{j=1,\dots,d}\}_{t \in \mathbb{Z}}$, where $\mathbf{X}_{t}$ follows a vector autoregressive model of order $p$, $\mathrm{VAR}(p)$, if
\begin{equation}
\label{vark}
\mathbf{X}_{t} = 
\boldsymbol{\Phi}_{1} \mathbf{X}_{t-1} + 
  \ldots + 
  \boldsymbol{\Phi}_{p} \mathbf{X}_{t-p} + 
  \mathbf{E}_{t}, \quad t \in \mathbb{Z},
\end{equation}
\noindent for some $d \times d$ matrices $\boldsymbol{\Phi}_{1}, \dots,\boldsymbol{\Phi}_{p}$ and a white noise series $\{\mathbf{E}_{t}\}_{t \in \mathbb{Z}} \sim \text{WN}(\mathbf{0}, \boldsymbol{\Sigma}_{\mathbf{E}})$ characterized by $\mathbb{E}(\mathbf{E}_{t})=0$ and $\mathbb{E}(\mathbf{E}_{t}\mathbf{E}^{'}_{s})=0$ for $s \neq t$. For simplicity we assume $\mathbf{X}_{t}$ is of zero mean, however, all developments that follow hold for models with a mean. Generally, a unique stationary solution to \eqref{vark} can be ensured by satisfying the stability condition given by $\text{det}( \boldsymbol{\Phi}(z)) \neq 0$, for $|z| \leq 1$, $z \in \mathbb{C}$, where  $\boldsymbol{\Phi}(z)=\mathbf{I}_{d} -\boldsymbol{\Phi}_{1}z - \dots - \boldsymbol{\Phi}_{p} z^{p}$. In addition, we can rewrite \eqref{vark} in more concise notation as
\begin{equation}
\label{bigvar}
\mathbf{Y}=\boldsymbol{\Phi}\mathbf{Z} + \mathbf{U},
\end{equation}
\noindent where $\mathbf{Y}$ is an $d \times (T-p)$ outcome matrix, $\boldsymbol{\Phi}$ is an $d \times (dp)$ transition matrix, $\mathbf{Z}$ is an $(dp) \times (T-p)$ design matrix, and $\mathbf{U}$ is an $d \times (T-p)$ matrix of process noise. In the remainder of this work we refer only to first order ($p=1$) VAR models with no mean structure, however, as in \eqref{bigvar} all arguments can be extended to handle mean structures and arbitrary lag orders without any loss of generality. \\
\subsubsection{OLS Estimation of the VAR Model}
For data from a single individual, ordinary least-squares (OLS) regression is commonly used to estimate \eqref{bigvar}. When $\boldsymbol{\Phi}$ is unrestricted the OLS estimates are asymptotically equivalent to estimates obtained using Generalized Least Squares \citep[GLS;][]{zellner1962}. Assuming  $\mathbf{U}_{t} \sim \mathcal{N}(\mathbf{0}, \boldsymbol{\Sigma}_{\boldsymbol{U_{t}}})$ are independent across $t$, the component-wise OLS estimates are Maximum Likelihood (ML) estimates \citep{lutkepohl2007}. Under some mild assumptions, these estimators are known to be asymptotically normal with explicit variance-covariance matrices.\\
\subsubsection{Penalized Estimation of the VAR Model}
A drawback of the unrestricted VAR model is its profligate parameterization \citep{sims1980}. Indeed, the number of VAR parameters grows quadratically with each component series included. As a result a large number of unknown coefficients must be estimated relative to the information available in the data, and with typical sample sizes, VAR estimates will lack precision and exhibit poor forecasting performance \citep{bruggemann2012}.\\
To address these dimensionality issues, much work has been devoted to reducing the number of non-trivial parameters in the VAR model space. One class of approaches accomplishes this by imposing restrictions on specific VAR coefficients (e.g. $\phi_{13}=\phi_{42}=0$). For example, \citet{lutkepohl2013} discussed top-down and bottom-up sequential search procedures for identifying parsimonious VAR models, while \citet{hsu2008} proposed using the least absolute shrinkage and selection operator \citep[Lasso;][]{tibshirani1996} to overcome the deficits of sequential specification searches. In terms of the properties of these approaches, consistency of standard Lasso estimation for single VAR models was established in the seminal paper by \cite{basu2015a,basu2015b}, extending results from the linear regression setting by \citet{loh2012a,loh2012b}. \\
\section{The multi-VAR Framework}
In this section we extend our discussion of the VAR model to the case where multivariate time series data is available from multiple subjects. We introduce the standard multi-VAR proposed by \cite{fisher2022} and discuss the general mechanics of the approach. Examples are provided illustrating these mechanics using the \texttt{multivar} package for R \citep{multivar}. We discuss some limitations of the standard implementation and present novel adaptive penalty weights designed to address these issues. Lastly we introduce a blocked-fold cross-validation approach for identifying the multi-VAR penalty parameters.\\
\subsection{Multiple Subject Vector Autoregressive Model}
With data from multiple individuals available we are now interested in estimating the transition matrices, $\boldsymbol{\Phi}^{1},\dots,\boldsymbol{\Phi}^{K}$ corresponding to $1,\dots,K$ individuals. The approach described herein relies on the following decomposition of $\boldsymbol{\Phi}$,
\begin{equation}
\label{decomp}
\boldsymbol{\Phi}^{k}= \boldsymbol{\Gamma}^{0}+ \boldsymbol{\Gamma}^{k},\quad k=1,\ldots,K,
\end{equation}
\noindent where $\boldsymbol{\Gamma}^{0} \in \mathbb{R}^{d \times d}$ corresponds to the common effects across $K$ individuals and $\boldsymbol{\Gamma}^{k} \in \mathbb{R}^{d \times d}$ corresponds to the effects unique to individual $k$. That is, each individual's transition matrix is the superposition of common effects shared by all individuals, and the unique effects specific to each individual. Notice there are no distributional assumptions placed on the common or unique effects, allowing for individual transition matrices to differ qualitatively and quantitatively across individuals. Specifically, we are interested in the case where $\boldsymbol{\Gamma}^{0}$ and $ \boldsymbol{\Gamma}^{k}$ are sparse.

\subsection{The Standard multi-VAR}

\cite{fisher2022} described one approach for solving \eqref{decomp} using a modified Lasso penalty, 
\begin{equation}
\label{multivar}
   \argmin_{\boldsymbol{\Gamma} = (\boldsymbol{\Gamma}^{0}, \boldsymbol{\Gamma}^{1},\dots,\boldsymbol{\Gamma}^{K})} 
   \frac{1}{N} \sum_{k=1}^{K}\|\mathbf{Y}^{k}-(\boldsymbol{\Gamma}^{0}+ \boldsymbol{\Gamma}^{k})\mathbf{Z}^{k}\|_{2}^{2}+
    \lambda_{1}\|\boldsymbol{\Gamma}^{0}\|_{1}+\sum_{k=1}^{K}\lambda_{2,k}\|\boldsymbol{\Gamma}^{k}\|_{1},
\end{equation}
where $\|\mathbf{A}\|_{1}$ denotes the $\ell_{1}$ norm of $\mathrm{vec}(\mathbf{A})$. Here we induce sparsity in the individual transition matrices, $ \boldsymbol{\Phi}^{k}$, through the decomposition of common and unique effects. Sparsity in the multi-VAR solution is governed by the penalty parameters $ \lambda_{1}$ and $\lambda_{2,k}$ chosen using cross-validation. Importantly, heterogeneity of the solution is also determined by the competition of the two penalty parameters. \\

Here it is helpful to consider three common situations. First, suppose individuals share little in common. In this case a sensible approach to estimating $K$ related VAR models would return essentially independent solutions. Specifically, for large enough values of $\lambda_{1}$, $\hat{\boldsymbol{\Gamma}}^{0}=\mathbf{0}$, and we obtain $\hat{\boldsymbol{\Phi}}^{k}=\hat{\boldsymbol{\Gamma}}^{k}$.  Second, suppose there was very little heterogeneity in a given process across individuals. Here, a sensible approach would pool the time series and estimate a single transition matrix. In this case, large enough values of $\lambda_{2,k}$ will lead to $\hat{\boldsymbol{\Gamma}}^{k}=\mathbf{0}$, and we will have $\hat{\boldsymbol{\Phi}}^{k}=\hat{\boldsymbol{\Gamma}}^{0}$. Third, if the heterogeneity in dynamics among the $K$ individuals is unknown in advance, and  both common and unique features are plausible, a sensible approach would return a model in which the common $\hat{\boldsymbol{\Gamma}}^{0}$ and unique effects $\hat{\boldsymbol{\Gamma}}^{k}$ are developed in accordance with how similar the $K$ individuals are.\\

\begin{figure}[h]
\caption{Toy Example of Heterogeneity Adjustments}
\label{fig:d}
\centering
\includegraphics[width=.9\textwidth]{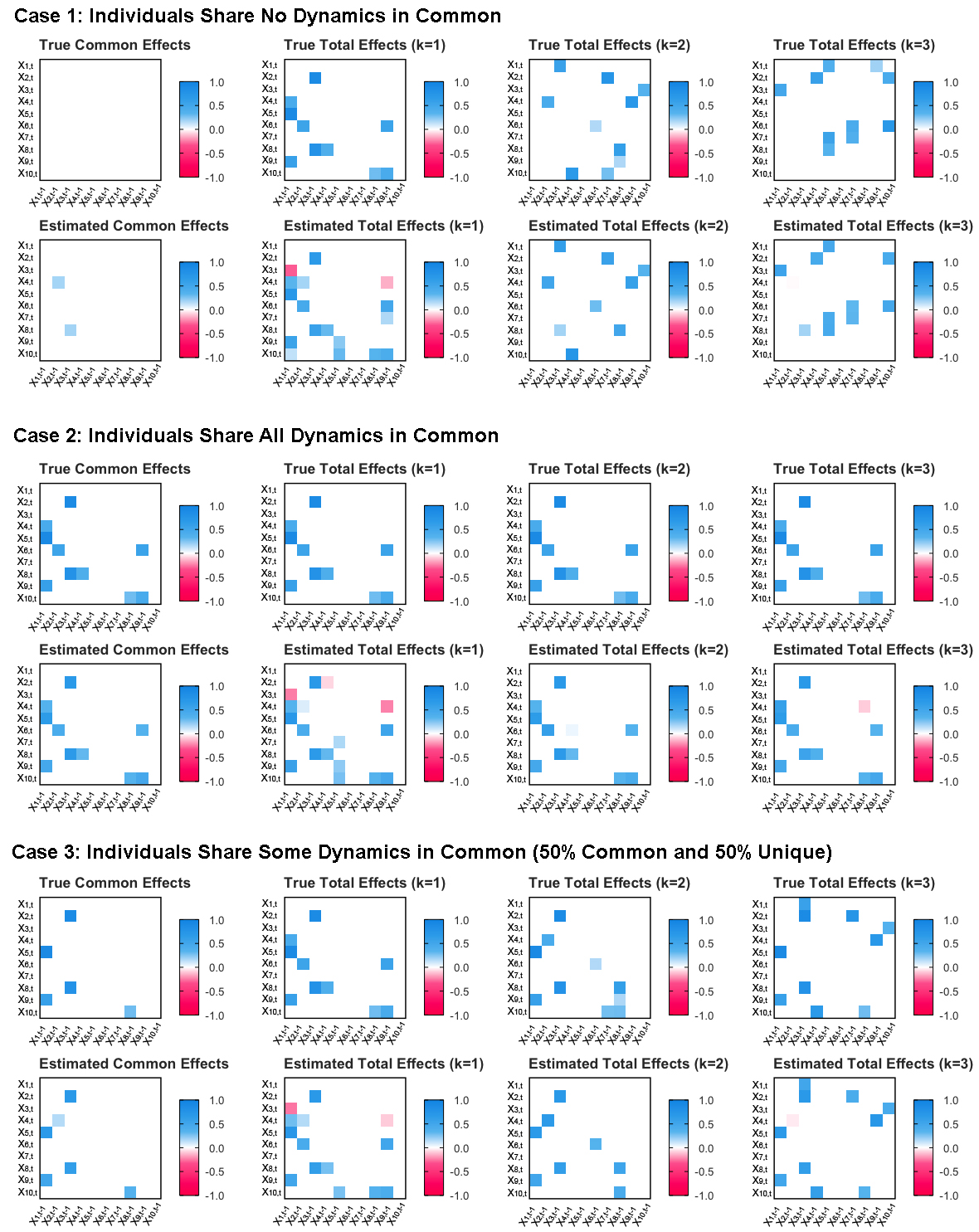}
\caption*{Data generating (true) and estimated transition matrices from multi-VAR are presented for three different cases, each exhibiting a different level of heterogeneity. For all cases, 10-variable first-order VAR models were generated for 3 individuals and 100 timepoints. In Case 1 data were generated such that all individuals exhibited unique dynamics, no nonzero elements were shared across individuals. In Case 2 data were generated such that all individuals showed the same pattern of nonzero elements in their transition matrices. In Case 3 data were generated such that 50\% of an individual's nonzero paths were common to all 3 individuals, and 50\% were entirely unique to each individual. Elements in the transition matrices represent parameter values whose magnitude is indicated by the legend.}
\end{figure}

It is also worth noting that just because a path is shared by a subset of individuals does not mean the estimated value of the corresponding parameter will be equal across individuals. In this way multi-VAR accommodates both qualitative and quantitative heterogeneity. See Figure \ref{fig:d} for a simulated toy example of these three settings (in the order of the figure, (1) heterogeneous features, (2) homogeneous features, and (3) heterogeneous and homogenous features) with $T=100$, $K=3$ and $d=10$. All models were estimated using the \texttt{multivar} R package \citep{multivar}. Code producing these results is available in the Supplementary Materials.\\

\subsection{The Adaptive multi-VAR}

It is worth noting the standard Lasso procedure is characterized by some important limitations, nicely discussed and summarized in a recent review  paper by \citet{freijeiro2022}. First, for consistent path selection by Lasso, the design matrix needs to satisfy additional assumptions such as the so-called irrepresentable condition  \citep{zhao2006}. These assumptions require the covariates not to be too strongly dependent, and are difficult to verify in practice. Second, even when the covariates are independent, Lasso is well-known to suffer from an excessive number of false positives. This is a delicate point having to do with the shrinkage procedure itself, thought to add pseudo-noise to the model (see section 2.1.3 in \citet{freijeiro2022}, and references therein). Moreover, bias can be a problem with $\ell_{1}$ penalization, primarily resulting from the shrinkage applied to coefficient estimates corresponding to the true signal \citep{buhlmann2011}. In certain cases, the bias associated with large estimates is equivalent to the penalty parameter used in soft-thresholding. We will revisit this final point in the following section.\\

\cite{zou2006} proposed the adaptive Lasso to overcome some of the limitations associated with the original Lasso procedure. The adaptive Lasso essentially replaces the $\ell_{1}$ penalty with a re-weighted version, where the weights are determined by some initial consistent estimate of the model parameters.  This weighting allows for differential penalization across the model coefficients, such that if an initial coefficient estimate is large, a smaller penalty is applied. The opposite holds for initial coefficient estimates that are small in magnitude, where a larger penalty is applied, and a more sparse solution is obtained. Remembering the bias in the $\ell_{1}$ penalty for large coefficients is often proportional to the penalty parameter, it is straightforward to see applying a smaller penalty to large coefficients will reduce the bias of the estimator. Likewise, it is also intuitive that applying larger penalties to coefficients with small initial estimates will reduce the number of false positives.\\

In addition to the standard approach, \cite{fisher2022} also proposed an adaptive version of the multi-VAR problem based on a between-person weighting scheme described by \cite{ollier2017}. The adaptive weights are built using a preliminary estimate $\tilde{\boldsymbol{\Phi}}^k = (\tilde\phi_{i,j}^{k})$ according to the following penalty function
\begin{equation}
\label{multivaradapt}
\mathcal{P}_{A} = 
    \lambda_{1} \sum_{i,j=1}^d \frac{1}{|\tilde{\phi}_{i,j,\textrm{median}}|^\alpha} |\Gamma_{i,j}^{0}| 
    + \sum_{k=1}^{K}\lambda_{2,k}\sum_{i,j=1}^d\frac{1}{|\tilde{\phi}_{i,j}^{k} - \tilde{\phi}_{i,j,\textrm{median}}|^{\alpha}}|\Gamma_{i,j}^{k}|
\end{equation}
where $\Gamma^{0}_{i,j}$ corresponds to $\{i,j\}^{th}$ element of the common effects matrix, $\boldsymbol{\Gamma}^{0}$, and  $\Gamma^{k}_{i,j}$ corresponds to the $\{i,j\}^{th}$ element of the unique effects matrix, $\boldsymbol{\Gamma}^{k}$. The divisors $|\tilde{\boldsymbol{\Phi}}_{median}|=(|\tilde{\phi}_{i,j,\textrm{median}}|)$ and $| \tilde{\boldsymbol{\Phi}}^k-\tilde{\boldsymbol{\Phi}}_{median}| = (|\tilde{\phi}_{i,j}^{k} - \tilde{\phi}_{i,j,\textrm{median}}|)$ are defined using the entrywise weighted medians of $(\tilde{\boldsymbol{\Phi}}^{1},\ldots,\tilde{\boldsymbol{\Phi}}^{K})$ and $\alpha \geq 1$. 

\subsection{Extending the multi-VAR Adaptive Weights}

In the regression setting, \citet{zhou2009} suggest any initial estimate $\tilde\theta$ of $\theta$ with a small bound on $||\tilde\theta - \theta||_{\infty}$. In practice, the choice of initial estimates can have a large impact on the bias and overall sparsity of the adaptive Lasso solution. \cite{fisher2022} used the unpenalized MLE of $\tilde{\boldsymbol{\Phi}}^k$  to construct the adaptive weights.  However, other initial estimators have been explored in the literature. For example, \cite{zou2006} suggested ridge regression and \cite{zhou2009}  suggested the Lasso as promising estimators for constructing the initial weights needed for the adaptive approach. Indeed the multi-VAR approach is even more sensitive to the choice of adaptive weights as adjustments are needed for coefficients not immediately estimable from the data (common and unique effects). For this reason it is important to look at how these alternative methods for weight construction impact the adaptive multi-VAR results.

\subsection{Cross-Validation and Penalty Parameter Selection}

The multi-VAR solution quality depends on the selection of the unknown penalty parameters $\lambda_{1}$ and $\lambda_{2, k}$, $k=1,\dots,K$. In the multi-VAR setting we construct a grid of plausible values for these parameters using the approach described in \citet{fisher2022}. To identify the optimal parameter values from a grid of plausible values we use cross-validation. \citet{fisher2022} proposed adapting a rolling-window cross-validation (RWCV) procedure for high-dimensional VAR models \citep{banbura2010, song2011} to the multiple-subject setting. However, this approach relies on 1-step ahead cross-validation and is computationally expensive. In the current work we propose a computationally efficient adaptation of traditional fold-based cross-validation using blocked sampling (BCV) to the multiple-subject time series context. Although the BCV approach does not preserve the temporal ordering of the component time series (a model can be tested on data that precedes chronologically the training data), it makes more efficient use of the available data compared to RWCV, where many timepoints are essentially removed from testing and training. Numerous authors have found BCV methods work well for single-subject autoregressive models and stationary time series data  \citep{bulteel2018a, bergmeir2012, bergmeir2014, bergmeir2018}.\\

\begin{figure}[h]
\caption{Examples of Blocked Cross-Validation for multi-VAR}
\label{fig:e}
\centering
\includegraphics[width=1\textwidth]{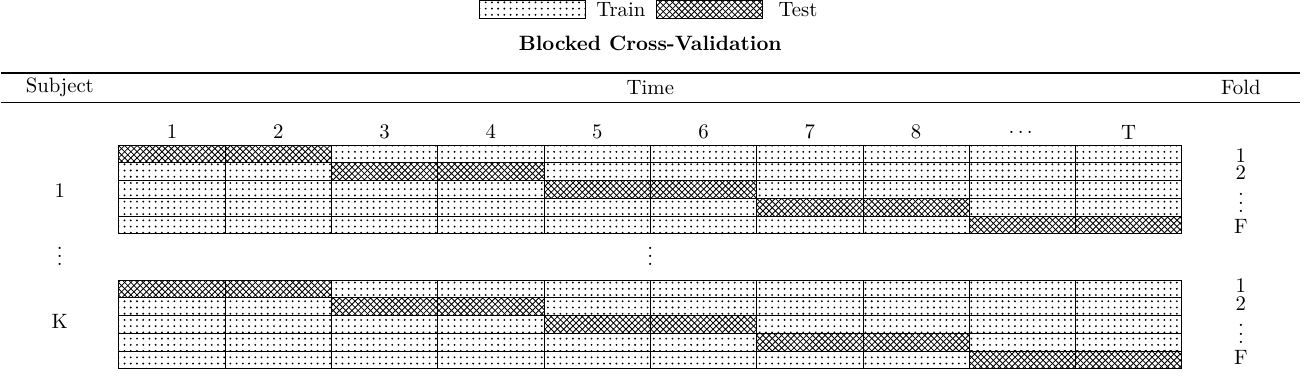}
\caption*{An example of blocked cross-validation for multiple-subject time series data. For any given fold, blocks marked Train are used to fit the model, and blocks marked Test are used to evaluate the prediction.}
\end{figure}

Figure \ref{fig:e} provides a visual depiction of the BCV approach. For each value of the $\lambda_1$ and $\lambda_{2,k}$ grid we perform the following sequence. First, we split each individual's times series into $F$ equivalently sized folds. Next, we remove one of the $1,\dots,F$ folds from each individual's multivariate time series (marked as the Test block in Figure \ref{fig:e}). Using all remaining folds (marked as the Train blocks in Figure \ref{fig:e}) we solve the problem in \eqref{multivaradapt} using the training blocks from each individual to obtain $\hat{\boldsymbol{\Gamma}}^{0}$ and $\hat{\boldsymbol{\Gamma}}^{k}$. Separately for each individual, these estimates are then used to forecast the testing block and obtain the MSFE. We continue in this fashion, removing each of the $F$ folds, forecasting the omitted test data, and calculating the error, at which time the forecast performance is aggregated across the $K$ individuals and $F$ folds for each combination of $\lambda_{1}$ and $\lambda_{2,k}$ as in
\begin{equation}
\text{MSFE}_{\lambda_{1},\lambda_{2,k}}= \frac{1}{K}
\sum^{K}_{k=1}\frac{1}{F}
\sum^{F}_{f=1}\| \hat{\mathbf{Y}}^{k}_{f}-\mathbf{Y}^{k}_{f} \|^{2}_{2},
\end{equation}
and the values of $\lambda_{1}$ and $\lambda_{2,k}$ which correspond to the smallest MSFE are chosen for the final model.

Although a thorough discussion of solving the multi-VAR optimization problem is beyond the scope of the current paper interested readers should see \citet{fisher2022}, describing a proximal gradient algorithm based on Fast Iterative Shrinkage-Thresholding Algorithm \citep[FISTA;][]{beck2009}. This approach is implemented in the \texttt{multivar} R package \citep{multivar}.

\section{Simulation Study}

To better understand how the initial adaptive weight estimators impact the overall performance of the adaptive multi-VAR we conducted a Monte Carlo simulation study. In doing so we also wanted to contextualize the performance of adaptive multi-VAR with respect to current approaches for decomposing group and individual-level effects in the VAR model (e.g. GIMME-VAR), and related $K=1$ approaches. The performance of $K=1$ approaches were of particular interest because they provide a benchmark for evaluating whether the recovery of individual-level model parameters is improved by sharing information across individuals. This question is especially interesting when considering different levels of heterogeneity, and the current simulations were designed to reflect a wider range of heterogeneous effects than those examined in \citet{fisher2022}. Finally, the simulations were broadly designed to provide additional context to results from the empirical example, an fMRI study assessing habitual control.  For this reason the number of variables and individuals was kept constant at values similar to the empirical data. However, time series length and other design features were varied to expand the utility of our simulations. In the remainder of this section, we provide an overview of the simulation design and results.

\subsection{Simulation Design}

\subsubsection{Approaches Considered}

In the simulation we compared different approaches for estimating individual-level transition matrices. For the adaptive multi-VAR we considered three different estimators for resolving the initial weights, $(\tilde{\boldsymbol{\Phi}}^{1},\ldots,\tilde{\boldsymbol{\Phi}}^{K})$, in \eqref{multivaradapt}: (1) maximum likelihood, (2) ridge regression and (3) Lasso. Outside of these initial weights the estimators were identical. To contextualize multi-VAR's performance we also included a number of comparison approaches. First, we considered a version of GIMME based on the VAR(1) model, GIMME-VAR. Second, we considered a number of estimators for the $K=1$ VAR model, where data from each individual is treated separately. These estimators represent some of the most common estimators for estimating $K=1$ VAR models: (1) unpenalized maximum likelihood where non-significant paths were set to zero using an $\alpha=0.05$ to achieve sparsity (ML Thresholded), (2) adaptive lasso with maximum likelihood weights, (3) adaptive lasso with ridge regression weights,  (4) adaptive Lasso with Lasso weights, and (5) a $K=1$ version of GIMME-VAR where only the individual-level model search is conducted (see the \texttt{indSEM()} procedure in the \texttt{gimme} R package \citep{gimme} documentation). \\

\subsubsection{Simulating Heterogeneous Time Series}

The proposed simulations were designed to produce a wider range of heterogeneity across individual dynamics than has previously been considered. \citet{fisher2022} also considered three heterogeneity conditions: in the low-heterogeneity condition $2/3$ of paths were common to all individuals, and $1/3$ of paths were completely unique to each individual; in the medium heterogeneity condition $1/2$ of each individual's paths were common and $1/2$ were unique; in the high-heterogeneity condition $1/3$ were common and $2/3$ were unique. This design produced considerably more heterogeneity than is typically considered in similar simulations, however, all paths were either unique to a single individual or shared by everyone. While this data generation approach is well-suited to multi-VAR (or GIMME-VAR), it is likely unrealistic for many processes. For example, it may be the case that certain paths are shared by some individuals and not others. In this simulation we sought to improve upon previous designs by generating data in a more realistic manner. Below we describe how data were generated for different heterogeneity conditions in the current simulation.\\
Consider the general multi-VAR setting, where the possibility of shared model structures among the $K$ individuals is allowed. In this setting, the degree of heterogeneity can vary, ranging from all, to only selected individuals, sharing common non-zero paths. To systematically investigate this idea we utilized the following procedure to quantify and generate heterogeneity when choosing paths to share across individuals. First, consider a multi-VAR model of order $p=1$, involving $d^2$ possible paths in $\boldsymbol{\Phi}^{k}$. Let $\pi_P = (\pi_{P,1},\ldots,\pi_{P,G})$ and $\pi_I = (\pi_{I,1},\ldots,\pi_{I,G})$ be vectors of proportions in $(0,1]$ associated with paths and individuals, respectively, such that $\pi_{P,1}+ \ldots+ \pi_{P,G}\leq 1$ and $\pi_{P,1}> \ldots> \pi_{P,G}$. For $g=1,\ldots,G$, $\pi_{P,g} = d_g/d^2$ (with integer $d_g$) refers to the proportion of paths that will be shared across $\pi_{I,g}=k_g/K$ proportion of individuals. The $d_g$ paths are assumed to be selected at random from those not chosen in previous steps $1,\ldots,g-1$, and the $k_g$ individuals are selected at random from all $K$ individuals for any $g$. As illustrated in the example below, $G$ represents the number of draws of individuals and paths considered within a given heterogeneity condition. In the current simulation we considered three heterogeneity conditions: (1) \emph{no heterogeneity} where $\pi_P = (1/4)$, $\pi_I = (1)$, and (2) \emph{low heterogeneity} where $\pi_P = (1/5,1/10,1/20)$, $\pi_I = (1, 2/3,1/3)$, and (3) \emph{high heterogeneity} where $\pi_P = (1/5,1/10,1/20)$, $\pi_I = (1/3, 2/3, 1)$.\\

Broadly speaking, in the \emph{no heterogeneity} condition, we have path proportions vector $\pi_P = 1/4$ and individual proportions vector $\pi_I = 1$. So, there is one sampling group consisting of $100\%$ of all individuals ($\pi_I = 1$), and $25\%$ of all paths are non-zero in this group.  Although the numeric values of the non-zero paths are distinct across individuals, the location of the non-zero paths are uniform. In the \emph{low heterogeneity} condition, $\pi_P = (1/5,1/10,1/20)$ and $\pi_I = (1, 2/3,1/3)$.  The first sampling group consists of $100\%$ of all individuals and has $20\%$ non-zero paths.  The second sampling group – which necessarily overlaps with the first group – consists of $2/3$ of all individuals and adds $10\%$ new non-zero paths that were previously zero.  Finally, the third group – again, which is a subset of the first group and may also overlap with the second group – consists of $1/3$ of all individuals and adds another $5\%$ new non-zero paths.  Just as with the \emph{no heterogeneity} condition, although the locations of non-zero paths may be shared across individuals, the specific non-zero values are distinct. In the \emph{high heterogeneity} condition, $\pi_P = (1/5,1/10,1/20)$ and $\pi_I = (1/3, 2/3, 1)$.  The \emph{high heterogeneity} condition is similar to the \emph{low heterogeneity} condition with the most important modification that only $5\%$ of all non-zero paths are shared across $100\%$ of all individuals, instead of $20\%$ of non-zero paths.  Importantly, the proportion of non-zero paths that is shared across individuals is much lower in the \emph{high heterogeneity} condition than in the \emph{low heterogeneity} condition. Additional clarification on our heterogeneity design is presented in the appendix, including a step-by-step description of the algorithm described above for a single heterogeneity condition.

\subsection{Additional Design Factors}

Data was generated according to time series length of  $30$, $50$, and $100$ for $15$ individuals and $10$ variables, in line with our empirical example. Although our empirical dataset had $100$ timepoints per individual, we also considered shorter time series lengths, as these are also quite common in the literature. Each dataset was then analyzed using the multi-VAR approach, GIMME-VAR and single-subject ($K=1$) methods. The multi-VAR approaches consisted of adaptive multi-VAR with either maximum likelihood, ridge, or lasso adaptive weights. The single-subject approaches consisted of unpenalized maximum likelihood and adaptive Lasso with either maximum likelihood, ridge, or lasso adaptive weights (labeled as $K=1$ approaches in the results). For all penalized approaches we used blocked cross-validation with 10 equal folds and mean-square forecast error (MSFE) calculated on each test fold to choose the hyper parameters. We also compared the blocked cross-validation (BCV) approach described above to the rolling-window cross validation (RWCV) approach described by \citet{fisher2022} for the adaptive multi-VAR approaches.  This was done to determine whether the proposed BCV method performs at least as well as the established method of RWCV. Additional details on these and other cross-validation procedures for time series can be found in \citet{cerqueira2020}. Finally, for each cell of the simulation design $750$ datasets were generated and analyzed.

\subsection{Outcome Measures} 
To compare the performance of the selected approaches we considered a number of metrics. To characterize variable selection performance, how well the different approaches recovered the data generating model, we used Matthew's correlation coefficient (MCC). MCC is a robust single-measure summary of path recovery that gives equal weight to positive and negative cases and incorporates both sensitivity and specificity in  its definition \citep{chicco2021matthews, chicco2020advantages}. MCC ranges from perfect disagreement ($MCC=-1$) to perfect agreement ($MCC=1$), with a value of $0$ indicating chance performance. For each multiple-subject dataset, the mean MCC is calculated as
\begin{align}
\label{mcc}
\text{Mean MCC} &= \frac{1}{K}\sum^{K}_{k=1}\left(\frac{
TP_{k} \times TN_{k} - FP_{k} \times FN_{k} 
}{(TP_{k}+FP_{k})(TP_{k}+FN_{k})(TN_{k}+FP_{k})(TN_{k}+FN_{k})}\right)
\end{align}
where the measure in \eqref{mcc} is averaged over $K$ individuals and \emph{TP} is the number of nonzero-valued parameters in the data generating model correctly recovered as nonzero-valued in the fitted model, \emph{TN} is the number of zero-valued parameters in the data generating model correctly recovered as zero-valued in the fitted model,  \emph{FP} is the number of zero-valued parameters in the data generating model incorrectly recovered as nonzero-valued in the fitted model, and \emph{FN} is the number of nonzero-valued parameters in the data generating model incorrectly recovered as zero-valued in the fitted model.\\
We also looked at the quality and variability of the estimated coefficients using absolute bias and root means square error (RMSE). For each multiple-subject dataset the mean absolute bias and RMSE were calculated as
\begin{align}
\text{Mean Absolute Bias} &= \frac{1}{K}\sum_{k=1}^K \frac{1}{d^2}\sum_{i,j=1}^d |\hat{{\phi}}^{k}_{i,j}-{\phi}^{k}_{i,j}|,\\
\text{Root Mean Square Error} &= \frac{1}{K}\sum_{k=1}^K \sqrt{\frac{1}{d^2}\sum_{i,j=1}^d (\hat{{\phi}}^{k}_{i,j}-{\phi}^{k}_{i,j})^2},
\end{align}
where ${\phi}^{k}_{i,j}$ and $\hat{\phi}^{k}_{i,j}$ are the $(k,i,j)^{th}$ elements of the data generating, and estimated transition matrices, respectively, for individual $k$ in a given design condition.\\

\subsection{Results}

\subsubsection{Cross-Validation Scheme}

Before comprehensively reviewing the simulations we note there was very little difference in the outcome measures of the adaptive multi-VAR estimators based on whether a rolling-window or blocked cross-validation procedure was used.  For those interested a direct comparison of the two approaches on all outcome measures is provided in the Appendix. However, as no meaningful differences were observed for the simulation conditions considered here all results discussed in the remainder of this section are based on the BCV procedure.\\

\subsubsection{Model Recovery}

\begin{figure}[h]
\caption{Simulation Results}
\label{fig:r}
\centering
\includegraphics[width=1\textwidth]{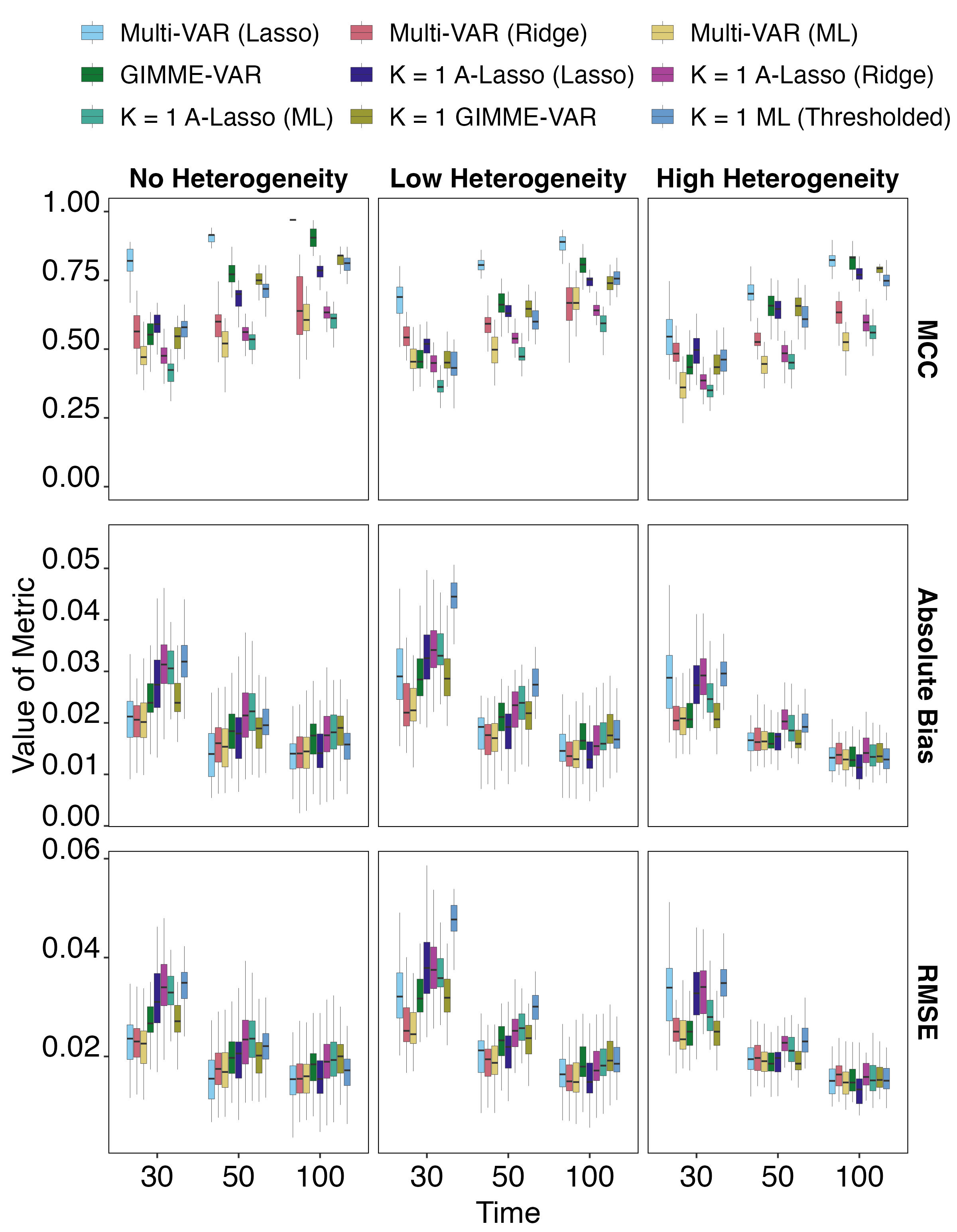}
\end{figure}

Path recovery was evaluated using MCC. Mean MCC values for each estimator, across simulations conditions, are provided in the first row of Figure \ref{fig:r}. In the No Heterogeneity condition, all individuals shared the same pattern of zero and nonzero elements (in their transition matrix), however, the data generating values for nonzero paths differed across individuals. In this condition, the multi-VAR approach with Lasso penalty weights performed best, with mean MCC values of $0.80$, $0.90$, and $0.95$ for the $T=30$, $T=50$, and $T=100$ conditions, respectively. It is worth noting these MCC values correspond to mean sensitivity values of $0.89$, $0.98$, and $0.98$, and mean specificity values of $0.94$, $0.96$, and $0.99$, for the three time series lengths. In the current context, sensitivity is a measure of the probability a path is recovered given it was in the data generating model, and specificity is the probability a path is not recovered given it was not in the data generating model. Formulas are calculating sensitivity and specificity are given in the Appendix.\\

In the Low Heterogeneity condition, individuals had more common paths than unique paths, and the data generating values of those paths varied across individuals. In this condition, the multi-VAR approach with Lasso penalty weights performed best, with mean MCC values of $0.67$, $0.80$, and $0.86$ for $T=30$, $T=50$, and $T=100$, respectively. These MCC values correspond to mean sensitivity values of $0.73$, $0.90$, and $0.95$, and mean specificity values of $0.92$, $0.93$, and $0.94$, for the three time series lengths.\\

In the High Heterogeneity condition, individuals had more unique paths than common paths, and again the data generating values of those paths varied across individuals. In this condition, the multi-VAR approach with Lasso penalty weights performed best for time series lengths of $T=30$, $T=50$, and comparably to the GIMME-VAR approach for the largest time series length, $T=100$. The multi-VAR with Lasso penalty weights resulted in mean MCC values of $0.54$, $0.71$, and $0.80$, while GIMME-VAR resulted in mean MCC values of $0.44$, $0.65$, and $0.81$, for $T=30$, $T=50$, and $T=100$, respectively. Consistent with results from the previous two conditions, increasing levels of heterogeneity tended to coincide with a decrement in path recovery, with those decrements primarily driven by decreases in sensitivity.\\

Overall, the multiple-subject approaches (multi-VAR and GIMME-VAR) tended to outperformed the $K=1$ approaches in path recovery performance, even when relatively few were paths shared by individuals in the sample.  For the different multi-VAR approaches, penalty weights computed with the Lasso were superior across all heterogeneity conditions. This result was driven by increases in specificity, where the Lasso weights provided a sparse initial estimate of individual-level transition matrices, $\tilde{\boldsymbol{\Phi}}^k, \dots, \tilde{\boldsymbol{\Phi}}^k$, which led to fewer false positives. In general, multi-VAR (Lasso) outperformed all other approaches in terms of path recovery, except for the high heterogeneity condition, at the largest time series length, where it performed similarly to GIMME-VAR. The performance differences between the adaptive multi-VAR (Lasso) and other approaches were most pronounced at smaller time series lengths and primarily driven by higher specificity values, meaning the adaptive multi-VAR (Lasso) was less likely to recover false positives than the other approaches. With increasing heterogeneity levels, all multiple-subject approaches tended to perform worse in terms of path recovery, becoming less sensitive.  Although sparsity levels (proportion of nonzero elements in an individual's transition matrix) were similar across heterogeneity conditions, they did vary, although not systematically. For example, mean sparsity was $0.20$, $0.28$, and $0.18$ in the No, Low and High Heterogeneity conditions, respectively. For this reason some caution is warranted in attributing performance differences to heterogeneity alone, where sparsity may also be a factor.

\subsubsection{Parameter Estimates}

We also considered two measures of parameter accuracy, mean absolute bias and RMSE. Both are measures of parameter accuracy, however, RMSE also provides information about the variance of the estimates. Overall, the relative pattern of the two measures was very similar, likely due to the overall sparsity of the recovered parameter matrices. For the No Heterogeneity condition, and across all time series lengths, mean absolute bias and RMSE tended to be lowest for the adaptive multi-VAR approaches, however, these differences were most pronounced at the shorter time series lengths.  For the Low Heterogeneity condition, the multi-VAR approaches with maximum likelihood and ridge penalty weights tended to exhibit the smallest absolute bias and RMSE, however, again, this was most visible at the smaller time series lengths. Lastly, for the High Heterogeneity condition, absolute bias and RMSE followed a similar pattern, except the differences between the multiple-subject approaches were less pronounced. It is also worth mentioning the $K=1$ adaptive Lasso approach performed as well or better than the multiple-subject approaches at time series lengths of $T=50$ and $T=100$.\\

In terms of absolute bias and RMSE, we again see the pattern of increasingly similar performance across all methods as heterogeneity, and time series length, increased. One interesting finding with regard to the absolute bias and RMSE is that among the adaptive multi-VAR approaches, the maximum likelihood penalty weights, and to a lesser extent ridge, tended to outperform the Lasso penalty weights. This was especially true at the shortest time series length of $T=30$.

\subsubsection{Recommendations}

Based on these simulations a few recommendations can be made in regard to using adaptive multi-VAR in practice. Most importantly, penalty weights based on the Lasso should be preferred in almost all circumstances for the adaptive multi-VAR procedure if model recovery is the goal. The adaptive multi-VAR approaches appeared to perform particularly well in the shorter time series lengths ($T=30$, $T=50$) compared to alternative approaches. Only in the high heterogeneity condition, at the shortest time series length of $T=30$, could an argument be made for using the maximum likelihood or ridge penalty weights, and even here the benefit is questionable. Another finding from these simulations is that as time series length and heterogeneity levels increase, the benefits of using adaptive multi-VAR (or any of the multiple-subject approaches that share information across individual datasets) decreases. Finally, we offer a word of caution in regard to extrapolating too far beyond the current simulation design. The current simulations were intended to mimic the empirical example, and only a small set of plausible data generating conditions were considered, namely VAR(1) models with $15$ individuals and $10$ variables. As such more work is needed to determine whether these conclusions generalize to additional contexts.

\subsection{Empirical Example}

Below we show how one might use adaptive multi-VAR to analyze multivariate time series data from multiple subjects. The goal of this analysis was to better understand the time-dependent relations underlying brain networks thought to be involved in habitual control. Empirical data from $12$ participants ($6$ adolescents, ages $15-17$ years, M\textsubscript{age} = 16, SD\textsubscript{age} = $0.89$; $6$ adults, ages $25-32$ years, M\textsubscript{age} = $28.5$, SD\textsubscript{age} = $2.43$) were included in the analyses. $10$ regions of interest (ROIs) were of interest as they represent 3 functional networks derived from the Power $264$ atlas \citep{power2011}. These networks include the cingulo-opercular task control network, the fronto-parietal task control network, and the default mode network (DMN) as they are broadly implicated in habitual control \citep{fox2005, uddin2009}. Time series data from each ROI were concatenated across participants to produce a single matrix with $10$ columns representing each ROI and $100$  timepoints.

\subsubsection{Procedure and Free-operant Task}

Participants completed a free-operant task \citep{tricomi2009} during the scan session where they learned to respond to fractal cues for food rewards. During the session participants were instructed to earn as many food rewards as possible, and to respond as often as they wanted. After each response, either a gray circle appeared for $50$ milliseconds, indicating no food reward, or a picture of a food reward appeared for $1$ second, indicating a reward was earned. If participants responded incorrectly, no display was shown. Food rewards were delivered on a $10$-second variable-interval schedule. After the session, participants were given the amount of food rewards they earned. The contingencies among fractal images, button presses, and food rewards remained consistent throughout the training sessions. Block order was pseudorandomized, with no fractal images repeating throughout the study.

\subsubsection{fMRI Data Acquisition and Preprocessing Procedures}

Images were acquired using a $3$-T Siemens MAGNETOM Prisma FIT Scanner with a $20$-channel head coil at the Social, Life, And Engineering Sciences Imaging Center at The Pennsylvania State University. High-resolution T1-weighted structural scans ($0.9 \times 0.9 \times 0.9$mm voxels) were collected. For the free-operant task, T2*-weighted gradient single-shot BOLD echo planar imaging (EPI) sequence functional images ($3 \times 3 \times 4$mm voxels, $TR = 2$s, $TE = 0.025$ms, flip angle = 70°, FoV = $250 \times 240$, slice gap = $0$mm) were collected.

Functional images were preprocessed using standard steps in Analysis of Functional NeuroImages (AFNI) software \citep{cox1996}. Structural and functional images were nonlinearly warped into MNI space. Images were corrected for slice timing effects. Translational and rotational head motion estimates were calculated. Images were aligned to the minimum outlier volume using a cost function (lpc+ZZ). Any movement that exceeded $0.3$mm compared to the previous volume, and TRs with greater than $5\%$ outlier intensity fraction were censored from deconvolution analysis. Smoothing of functional images was accomplished by applying a Gaussian filter set at 4.0mm full-width at half maximum. We then used AFNIs 3dDeconvolve for deconvolution analysis. Motion estimates, their derivatives, and a fourth-order polynomial function to remove scanner drift during scans, were included as covariates of no interest. Importantly, no task specific regressors were included in the model. The residual time series after deconvolution from each participants was used for subsequent analyses.

\subsubsection{Results}

Data were analyzed using the \texttt{multivar} package \citet{fisher2022} with the Lasso-based adaptive penalty weights. These weights were chosen because they resulted in the best path recovery performance for time series length of $T=100$ in the simulations. Furthermore, the Lasso-based penalty weights performed similarly to the other weights at $T=100$ in terms of bias and RMSE.

\begin{figure}[h]
\caption{Estimated Dynamics During Free-Operant Task}
\label{fig:z}
\centering
\includegraphics[width=1\textwidth]{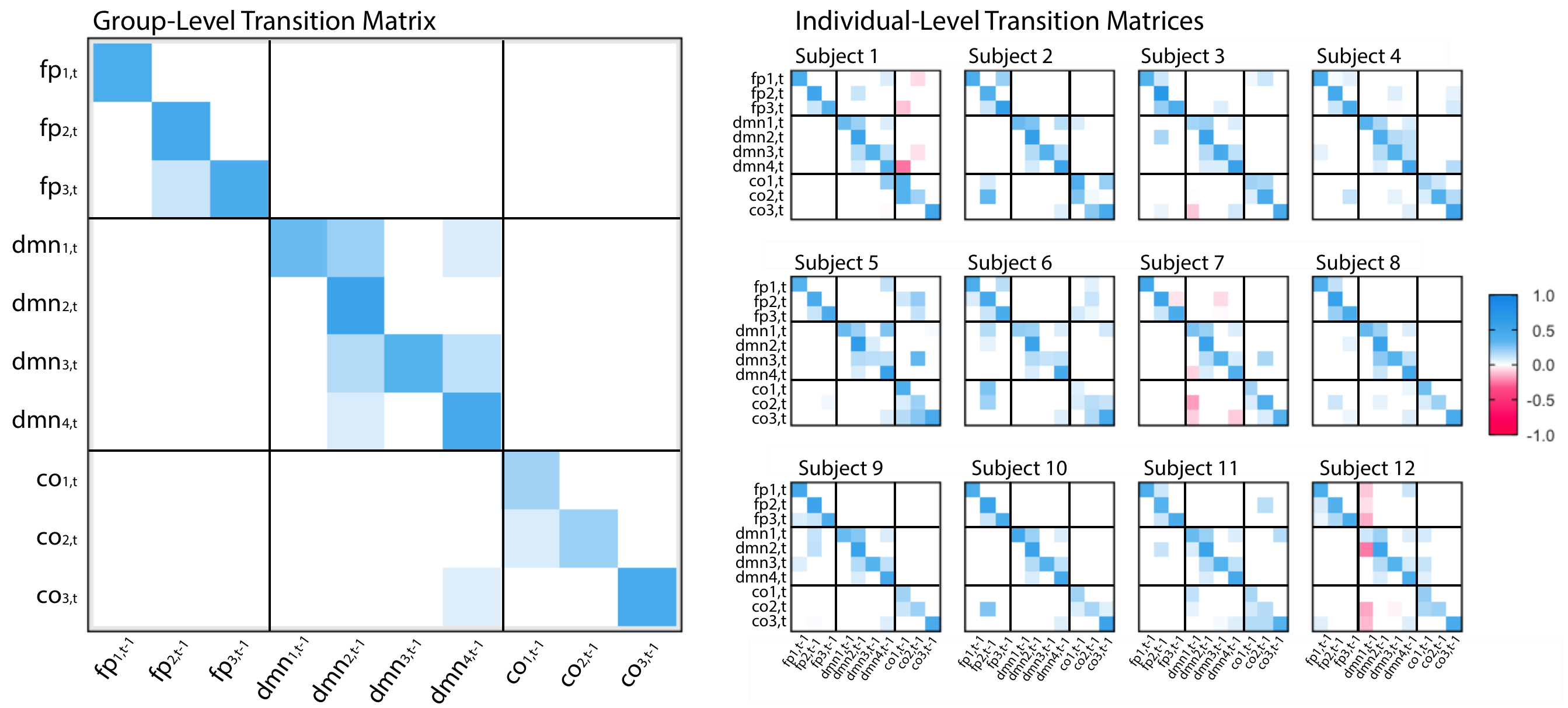}
\caption*{Note. Variables beginning with fp indicate regions of interest (ROIs) in the fronto-parietal network, variables beginning with dmn indicate ROIs in the default mode network, and variables beginning with co indicate ROIs in the cingulo-opercular network.}
\end{figure}

In looking at Figure \ref{fig:z} we can see the common and individual-level transition matrices for all 12 subjects. The common effects matrix shows a general set of dynamics recovered at the group-level. From the common effects it is clear lead-lag relations are stronger within each network, compared to the dependence between ROIs from different networks. Furthermore, average autoregressive path values were similar in magnitude between the fronto-parietal network ($M = 0.44$) and default mode network ($M = 0.42$), compared to smaller average autoregressive path values in the cingulo-opercular network ($M = 0.28$). This suggests the cingulo-opercular network may be better regulated during the prescribed learning tasks, compared to the other two networks.

The individual-level plots from the adaptive multi-VAR analysis are shown in Figure \ref{fig:z}. The unique patterns of brain connectivity illustrated in these plots during the free operant task may suggest differential susceptibility to form habits.  For example, a study from \citet{wang2022changes} found that changes in connectivity among nodes in the DMN and cingulo-opercular network were associated with individual differences in habit strength. Additionally, increased functional connectivity of the frontoparietal network has been implicated in obsessive compulsive disorder \citep{stern2012resting}, anorexia nervosa \citep{boehm2014increased}, and substance abuse \citep{zilverstand2020dual}. Thus, between-person heterogeneity in the qualitative structure of networks may serve as one indicator of maladaptive habit formation. Overall, these exploratory results suggest some interesting directions for future work looking at the time-dependence among networks implicated in habitual control.

\section{Discussion}

In a very real sense the study of temporal processes forces us to acknowledge the variation inherent to many psychological processes. Unfortunately, in the name of tradition and convenience, this variability is often treated as a vagary of generalizable research. John Nesselroade, reflecting on the use of time series models for characterizing psychological phenomena, and referencing Peter Molenaar's \emph{Manifesto on Psychology as Idiographic Science}, once wrote "the place of the individual in a science of behavior aspiring to establish general lawful relationships is somewhat ambiguous" \citep[][p. 249]{nesselroade2007} Over the last 50 years there is perhaps none who have done more to resolve this ambiguity than Peter Molenaar. Even now, how best to temporally synthesize heterogeneous dynamics across individuals remains an important question for a generalizable science of behavior. This is even more true when one is concerned with processes exhibiting qualitative and quantitative heterogeneity, where individuals may differ in the magnitude and overall pattern of time-dependent dynamics. \\

In this work we extended the adaptive multi-VAR approach proposed in \citet{fisher2022} to include new penalty weights and cross-validation schemes, the former of which dramatically improved path recovery across a variety of commonly encountered data scenarios. Performance of these new weights was evaluated in a small simulation study, and the multi-VAR approach with Lasso weights outperformed alternatives across the majority of simulation conditions. Most remarkably, multi-VAR outperformed $K=1$ approaches in path recovery, even when relatively few paths were shared by individuals in the sample. Simulation results also showed as heterogeneity levels and time series lengths increased, the benefits of multi-VAR over $K=1$ approaches diminished. Although these limited results appear promising a number of important questions remain unaddressed. Future work should consider comparing the multi-VAR approaches to alternative methods for characterizing heterogeneity in time series models, such as mixture modeling and flexible multilevel approaches. Additional work is also needed to disentangle sparsity and heterogeneity in proposing and evaluating models.

Practically speaking, methods for handling missing data within the multi-VAR framework are needed. Recent work from \cite{ji2018} considered multiple imputation (MI) approaches for handling missing data in multivariate time-series models, however, how best to integrate these approaches with penalized estimation is an open question. Critical to this endeavor is how to aggregate over imputation iterations when different sets of variables are retained for each iteration. Another open question relates to the manner in which the current multi-VAR implementation induces sparsity in individual-level transition matrices. In the current procedure, the penalization is applied to the common effects and unique transition matrices, $\boldsymbol{\Gamma}^{0}$ and $\boldsymbol{\Gamma}^{k}$, respectively. The penalty is not directly applied to the sum of these matrices (e.g. $\boldsymbol{\Phi}^{k}=\boldsymbol{\Gamma}^{0} + \boldsymbol{\Gamma}^{k}$).  Thus, sparsity in $\boldsymbol{\Phi}^{k}$ is encouraged indirectly, through this sum.

In certain situations this can result in small false positive values when an individual does not have a group-path in their final model. Lastly, the multi-VAR framework should be expanded to include additional time-series models, beyond the standard VAR (e.g. structural, graphical and time-varying VAR). Work addressing these important areas of development is currently underway. It is our hope the development of multi-VAR and similar approaches will provide researchers with tools for flexibly addressing between-person heterogeneity in time series dynamics. 

\newpage

\section{Appendix}

\subsection{Demo Code from \emph{Synthesizing Heterogeneous Dynamics} Section}

Below we include the code used to generate Figure \ref{fig:d}.

\begin{lstlisting}
(*@set@*).seed(1234)
sim  <- multivar::multivar_sim(
  k = 3,   # number of individuals
  d = 10,  # number of variables
  n = 100, # number of timepoints
  prop_fill_com = 0, # proportion of common that are common
  prop_fill_ind = 0.1, # proportion of paths that are unique
  lb = 0.1,  # lower bound on coefficient magnitude
  ub = 0.9,  # upper bound on coefficient magnitude
  sigma = diag(1,10) # noise
)

model <- multivar::constructModel(data = sim$data)
fit <- multivar::cv.multivar(model)


gridExtra::grid.arrange(
  plot_transition_mat(sim$mat_com, title = "True Common Effects"), 
  plot_transition_mat(sim$mat_ind_final[[1]], title = "True Total Effects (k=1)"), 
  plot_transition_mat(sim$mat_ind_final[[2]], title = "True Total Effects (k=2)"), 
  plot_transition_mat(sim$mat_ind_final[[3]], title = "True Total Effects (k=3)"), 
  plot_transition_mat(fit$mats$common, title = "Estimated Common Effects"), 
  plot_transition_mat(fit$mats$total[[1]], title = "Estimated Total Effects (k=1)"), 
  plot_transition_mat(fit$mats$total[[2]], title = "Estimated Total Effects (k=2)"), 
  plot_transition_mat(fit$mats$total[[3]], title = "Estimated Total Effects (k=3)"), 
  ncol = 4, nrow = 2
)

(*@set@*).seed(1234)
sim  <- multivar::multivar_sim(
  k = 3,   # number of individuals
  d = 10,  # number of variables
  n = 100, # number of timepoints
  prop_fill_com = 0.1, # proportion of common that are common
  prop_fill_ind = 0, # proportion of paths that are unique
  lb = 0.1,  # lower bound on coefficient magnitude
  ub = 0.9,  # upper bound on coefficient magnitude
  sigma = diag(1,10) # noise,
)

model <- multivar::constructModel(data = sim$data)
fit <- multivar::cv.multivar(model)


gridExtra::grid.arrange(
  plot_transition_mat(sim$mat_com, title = "True Common Effects"), 
  plot_transition_mat(sim$mat_ind_final[[1]], title = "True Total Effects (k=1)"), 
  plot_transition_mat(sim$mat_ind_final[[2]], title = "True Total Effects (k=2)"), 
  plot_transition_mat(sim$mat_ind_final[[3]], title = "True Total Effects (k=3)"), 
  plot_transition_mat(fit$mats$common, title = "Estimated Common Effects"), 
  plot_transition_mat(fit$mats$total[[1]], title = "Estimated Total Effects (k=1)"), 
  plot_transition_mat(fit$mats$total[[2]], title = "Estimated Total Effects (k=2)"), 
  plot_transition_mat(fit$mats$total[[3]], title = "Estimated Total Effects (k=3)"), 
  ncol = 4, nrow = 2
)

(*@set@*).seed(1234)
sim  <- multivar::multivar_sim(
  k = 3,   # number of individuals
  d = 10,  # number of variables
  n = 100, # number of timepoints
  prop_fill_com = 0.05, # proportion of common that are common
  prop_fill_ind = 0.05, # proportion of paths that are unique
  lb = 0.1,  # lower bound on coefficient magnitude
  ub = 0.9,  # upper bound on coefficient magnitude
  sigma = diag(1,10) # noise,
)
model <- multivar::constructModel(data = sim$data)
fit <- multivar::cv.multivar(model)

gridExtra::grid.arrange(
  plot_transition_mat(sim$mat_com, title = "True Common Effects"), 
  plot_transition_mat(sim$mat_ind_final[[1]], title = "True Total Effects (k=1)"), 
  plot_transition_mat(sim$mat_ind_final[[2]], title = "True Total Effects (k=2)"), 
  plot_transition_mat(sim$mat_ind_final[[3]], title = "True Total Effects (k=3)"), 
  plot_transition_mat(fit$mats$common, title = "Estimated Common Effects"), 
  plot_transition_mat(fit$mats$total[[1]], title = "Estimated Total Effects (k=1)"), 
  plot_transition_mat(fit$mats$total[[2]], title = "Estimated Total Effects (k=2)"), 
  plot_transition_mat(fit$mats$total[[3]], title = "Estimated Total Effects (k=3)"), 
  ncol = 4, nrow = 2
)
\end{lstlisting}

\subsection{Generating Heterogenous Dynamics}

To better illustrate the data generation algorithm presented earlier it may be helpful to consider a single condition in more detail. Let us supposed a 10-dimensional VAR(1) model with 15 individuals. Let us also consider the \emph{high heterogeneity} condition where we have path proportions vector, $\pi_P = (1/5,1/10,1/20)$, and individual proportions vector, $\pi_I = (1/3, 2/3, 1)$. We will simultaneously iterate through these two vectors in the following manner to construct the individual transition matrices a given instance of the  \emph{high heterogeneity} condition.\\
For iteration 1 we begin by taking the first element from the path proportions vector, $\pi_{P,1}=1/5$. This number indicates the proportion of nonzero paths that will be shared by the proportion of individuals in  $\pi_{I,1}$.  As we have $100$ possible nonzero paths in an arbitrary $10 \times 10$ transition matrix in this example, we randomly select the location of $20$ nonzero elements ($\pi_{P,1} = 1/5 = 20/100$). We now assign those nonzero paths to $\pi_{I,1}=1/3$ of our sample. As there are $15$ individuals in our current example, we randomly select $5$ individuals ($\pi_{I,1} = 1/3 = 5/15$) and assign those $5$ individuals to the same $20$ nonzero paths. Note, although the location of these $20$ non-zero paths are uniform across the $5$ individuals selected, the numeric values of the non-zero paths are distinct. \\
In iteration 2 we take the second element from the path proportions vector, $\pi_{P,2}=1/10$ and the second element from the individual proportions vector $\pi_{I,2}=2/3$. Using these proportions we randomly select $10$ new nonzero paths that were not previously selected in iteration 1 ($\pi_{P,2} = 1/10 = 10/100$).  We then assign these nonzero paths to $10$ randomly selected individuals ($\pi_{I,2} = 2/3 = 10/15$). Note, the randomly selected individuals from iteration 2 may overlap with the individuals selected in iteration 1. \\
In iteration 3 we take the third and final element from the path proportions vector, $\pi_{P,3}=1/20$ and the second element from the individual proportions vector $\pi_{I,3}=1$. Using these proportions we randomly  select $5$ new nonzero paths ($\pi_{P,3} = 1/20 = 5/100$) that were not previously selected in iteration 1 or 2, and assign them to all $15$ individuals in the sample ($\pi_{I,3} = 1 = 15/15$). Again we note that although the location of these non-zero paths are uniform across the individuals selected, the numeric values of those non-zero paths are distinct across individuals. \\
Finally, for all individuals separately, choose parameter values at random from the uniform distribution $U(0.1,0.9)$ for all nonzero paths identified in the previous iterations and, if needed,  re-scale the resulting transition matrices, $\boldsymbol{\Phi}^{k}$ to ensure a stable solution. Using their respective path and individual proportion vectors this exact procedure is then followed to construct the data generating matrices for each replication of \emph{no} and \emph{low heterogeneity} conditions.

\subsection{Sensitivity and Specificity}

In the paper we occasionally reference sensitivity and specificity when it provides additional insight into path recovery performance. Sensitivity and specificity are measures of the probability a path is recovered, given it was in the data generating model, and the probability a path is not recovered, given it was not in the data generating model, respectively. Here we show how these measures were calculated, 
\begin{align}
\label{sensitivity}
\text{Mean Sensitivity} &= \frac{1}{K}\sum^{K}_{k=1}\left(\frac{TP_{k}}{TP_{k}+FN_{k}}\right),\\
\label{specificity}
\text{Mean Specificity} &= \frac{1}{K}\sum^{K}_{k=1}\left(\frac{TN_{k}}{TN_{k}+FP_{k}}\right),
\end{align}
where aggregate measures were obtained by averaging across Monte Carlo iterations. Here, as in the calculation of MCC above, \emph{TP} is the number of nonzero-valued parameters in the data generating model correctly recovered as nonzero-valued in the fitted model, \emph{TN} is the number of zero-valued parameters in the data generating model correctly recovered as zero-valued in the fitted model,  \emph{FP} is the number of zero-valued parameters in the data generating model incorrectly recovered as nonzero-valued in the fitted model, and \emph{FN} is the number of nonzero-valued parameters in the data generating model incorrectly recovered as zero-valued in the fitted model.\\

\subsection{Comparison of Cross-Validation Schemes}

\begin{figure}[h]
\caption{Simulation-based Comparison of Cross-Validation Schemes}
\label{fig:j}
\centering
\includegraphics[width=1\textwidth]{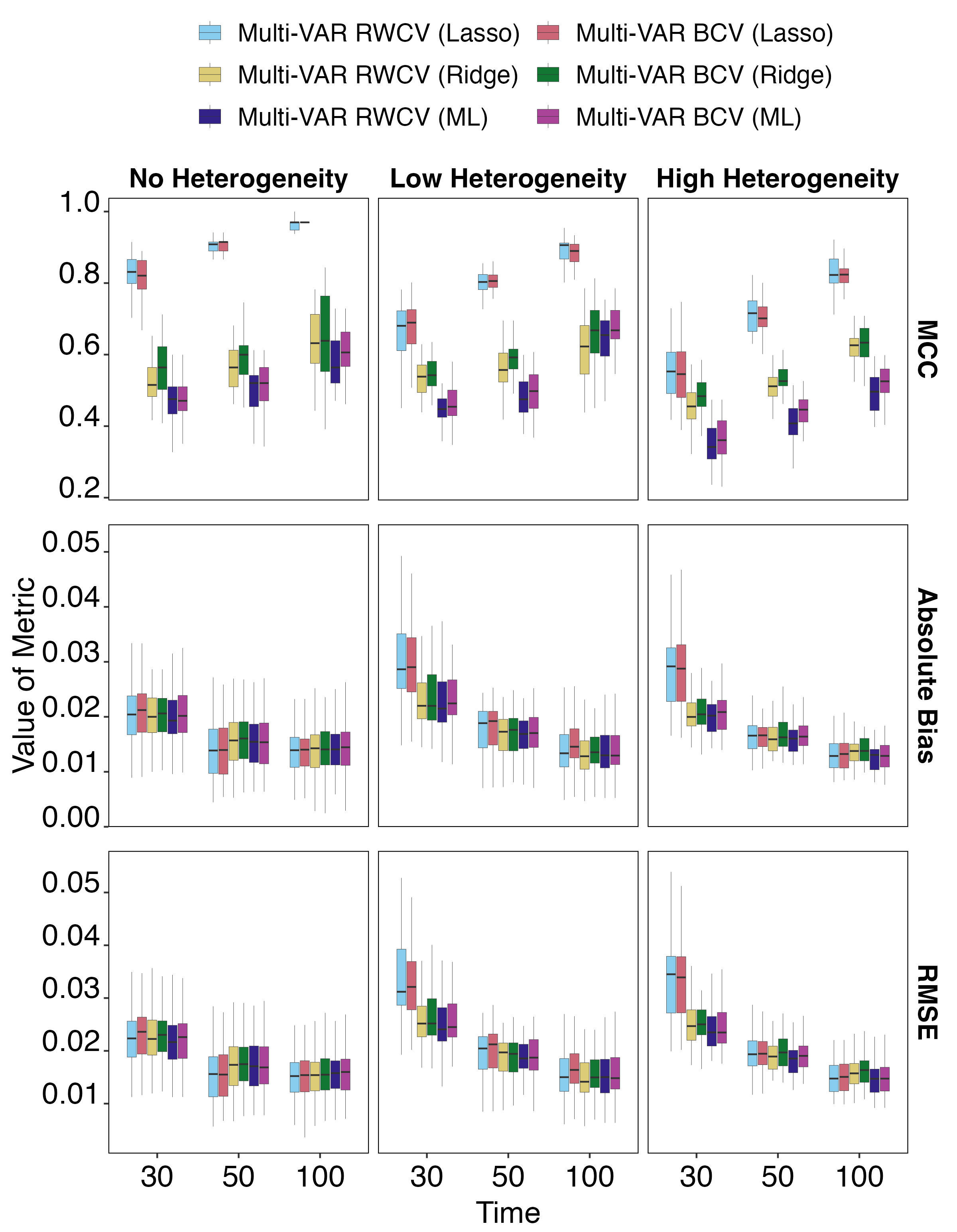}
\caption*{Note.  Adaptive multi-VAR approaches labeled BCV used blocked cross-validation and approaches labeled RWCV used rolling-window cross-validation.}
\end{figure}

Simulation results for the adaptive multi-VAR models using both the block and rolling-window cross-validation approaches are provided in Figure \ref{fig:j}. No discernible pattern of differences emerged across the two cross-validation approaches for the set of simulation conditions considered.

\begin{singlespace}
\bibliography{multivar_paper}

\begin{thebibliography}{}

\bibitem[Arizmendi et~al., 2021]{arizmendi2021}
Arizmendi, C., Gates, K., Fredrickson, B., and Wright, A. (2021).
\newblock Specifying exogeneity and bilinear effects in data-driven model
  searches.
\newblock {\em Behavior Research Methods}, 53(3):1276--1288.

\bibitem[Ba{\'n}bura et~al., 2010]{banbura2010}
Ba{\'n}bura, M., Giannone, D., and Reichlin, L. (2010).
\newblock Large {{Bayesian}} vector auto regressions.
\newblock {\em Journal of Applied Econometrics}, 25(1):71--92.

\bibitem[Basu and Michailidis, 2015a]{basu2015a}
Basu, S. and Michailidis, G. (2015a).
\newblock Regularized estimation in sparse high-dimensional time series models.
\newblock {\em Annals of Statistics}, 43(4):1535--1567.

\bibitem[Basu and Michailidis, 2015b]{basu2015b}
Basu, S. and Michailidis, G. (2015b).
\newblock Supplement to "{R}egularized estimation in sparse high-dimensional
  time series models".
\newblock {\em Annals of Statistics}.

\bibitem[Beck and Teboulle, 2009]{beck2009}
Beck, A. and Teboulle, M. (2009).
\newblock A {{Fast Iterative Shrinkage}}-{{Thresholding Algorithm}} for
  {{Linear Inverse Problems}}.
\newblock {\em SIAM Journal on Imaging Sciences}, 2(1):183--202.

\bibitem[Beltz and Molenaar, 2016]{beltz2016}
Beltz, A.~M. and Molenaar, P.~C. (2016).
\newblock Dealing with multiple solutions in structural vector autoregressive
  models.
\newblock {\em Multivariate behavioral research}, 51(2-3):357--373.

\bibitem[Bergmeir and Ben{\'i}tez, 2012]{bergmeir2012}
Bergmeir, C. and Ben{\'i}tez, J.~M. (2012).
\newblock On the use of cross-validation for time series predictor evaluation.
\newblock {\em Information Sciences: an International Journal}, 191:192--213.

\bibitem[Bergmeir et~al., 2014]{bergmeir2014}
Bergmeir, C., Costantini, M., and Ben{\'i}tez, J.~M. (2014).
\newblock On the usefulness of cross-validation for directional forecast
  evaluation.
\newblock {\em Computational Statistics \& Data Analysis}, 76:132--143.

\bibitem[Bergmeir et~al., 2018]{bergmeir2018}
Bergmeir, C., Hyndman, R.~J., and Koo, B. (2018).
\newblock A note on the validity of cross-validation for evaluating
  autoregressive time series prediction.
\newblock {\em Computational Statistics \& Data Analysis}, 120:70--83.

\bibitem[Boehm et~al., 2014]{boehm2014increased}
Boehm, I., Geisler, D., King, J.~A., Ritschel, F., Seidel, M., Deza~Araujo, Y.,
  Petermann, J., Lohmeier, H., Weiss, J., Walter, M., et~al. (2014).
\newblock Increased resting state functional connectivity in the
  fronto-parietal and default mode network in anorexia nervosa.
\newblock {\em Frontiers in behavioral neuroscience}, 8:346.

\bibitem[Bringmann et~al., 2013]{bringmann2013a}
Bringmann, L.~F., Vissers, N., Wichers, M., Geschwind, N., Kuppens, P.,
  Peeters, F., Borsboom, D., and Tuerlinckx, F. (2013).
\newblock A {{Network Approach}} to {{Psychopathology}}: {{New Insights}} into
  {{Clinical Longitudinal Data}}.
\newblock {\em PLoS ONE}, 8(4).

\bibitem[Br{\"u}ggemann, 2012]{bruggemann2012}
Br{\"u}ggemann, R. (2012).
\newblock {\em Model reduction methods for vector autoregressive processes},
  volume 536.
\newblock Springer Science \& Business Media.

\bibitem[Bryan et~al., 2021]{bryan2021}
Bryan, C.~J., Tipton, E., and Yeager, D.~S. (2021).
\newblock Behavioural science is unlikely to change the world without a
  heterogeneity revolution.
\newblock {\em Nature human behaviour}, 5(8):980--989.

\bibitem[B{\"u}hlmann and Van De~Geer, 2011]{buhlmann2011}
B{\"u}hlmann, P. and Van De~Geer, S. (2011).
\newblock {\em Statistics for high-dimensional data: methods, theory and
  applications}.
\newblock Springer Science \& Business Media.

\bibitem[Bulteel et~al., 2018]{bulteel2018a}
Bulteel, K., Mestdagh, M., Tuerlinckx, F., and Ceulemans, E. (2018).
\newblock Var (1) based models do not always outpredict ar (1) models in
  typical psychological applications.
\newblock {\em Psychological methods}, 23(4):740.

\bibitem[Bulteel et~al., 2016]{bulteel2016}
Bulteel, K., Tuerlinckx, F., Brose, A., and Ceulemans, E. (2016).
\newblock Clustering vector autoregressive models: Capturing qualitative
  differences in within-person dynamics.
\newblock {\em Frontiers in Psychology}, 7:1540.

\bibitem[Cerqueira et~al., 2020]{cerqueira2020}
Cerqueira, V., Torgo, L., and Mozeti{\v c}, I. (2020).
\newblock Evaluating time series forecasting models: An empirical study on
  performance estimation methods.
\newblock {\em Machine Learning}, 109(11):1997--2028.

\bibitem[Chicco and Jurman, 2020]{chicco2020advantages}
Chicco, D. and Jurman, G. (2020).
\newblock The advantages of the matthews correlation coefficient (mcc) over f1
  score and accuracy in binary classification evaluation.
\newblock {\em BMC genomics}, 21:1--13.

\bibitem[Chicco et~al., 2021]{chicco2021matthews}
Chicco, D., T{\"o}tsch, N., and Jurman, G. (2021).
\newblock The matthews correlation coefficient (mcc) is more reliable than
  balanced accuracy, bookmaker informedness, and markedness in two-class
  confusion matrix evaluation.
\newblock {\em BioData mining}, 14(1):1--22.

\bibitem[Cox, 1996]{cox1996}
Cox, R.~W. (1996).
\newblock Afni: software for analysis and visualization of functional magnetic
  resonance neuroimages.
\newblock {\em Computers and Biomedical research}, 29(3):162--173.

\bibitem[Duffy et~al., 2021]{duffy2021}
Duffy, K.~A., Fisher, Z.~F., Arizmendi, C.~A., Molenaar, P.~C., Hopfinger, J.,
  Cohen, J.~R., Beltz, A.~M., Lindquist, M.~A., Hallquist, M.~N., and Gates,
  K.~M. (2021).
\newblock Detecting task-dependent functional connectivity in group iterative
  multiple model estimation with person-specific hemodynamic response
  functions.
\newblock {\em Brain connectivity}, 11(6):418--429.

\bibitem[Epskamp et~al., 2018]{epskamp2018a}
Epskamp, S., Waldorp, L.~J., M{\~o}ttus, R., and Borsboom, D. (2018).
\newblock The {{Gaussian Graphical Model}} in {{Cross}}-{{Sectional}} and
  {{Time}}-{{Series Data}}.
\newblock {\em Multivariate Behavioral Research}, 53(4):453--480.

\bibitem[Fisher et~al., 2021]{multivar}
Fisher, Z., Kim, Y., and Pipiras, V. (2021).
\newblock {\em multivar: Penalized Estimation of Multiple-Subject Vector
  Autoregressive (multi-VAR) Models}.
\newblock R package version 1.1.0.

\bibitem[Fisher et~al., 2022]{fisher2022}
Fisher, Z.~F., Kim, Y., Fredrickson, B.~L., and Pipiras, V. (2022).
\newblock Penalized estimation and forecasting of multiple subject intensive
  longitudinal data.
\newblock {\em Psychometrika}, pages 1--29.

\bibitem[Foster and Beltz, 2021]{foster2021}
Foster, K.~T. and Beltz, A.~M. (2021).
\newblock Heterogeneity in affective complexity among men and women.
\newblock {\em Emotion}.

\bibitem[Fox et~al., 2005]{fox2005}
Fox, M.~D., Snyder, A.~Z., Vincent, J.~L., Corbetta, M., Van~Essen, D.~C., and
  Raichle, M.~E. (2005).
\newblock The human brain is intrinsically organized into dynamic,
  anticorrelated functional networks.
\newblock {\em Proceedings of the National Academy of Sciences},
  102(27):9673--9678.

\bibitem[Freijeiro-Gonz{\'a}lez et~al., 2022]{freijeiro2022}
Freijeiro-Gonz{\'a}lez, L., Febrero-Bande, M., and Gonz{\'a}lez-Manteiga, W.
  (2022).
\newblock A critical review of lasso and its derivatives for variable selection
  under dependence among covariates.
\newblock {\em International Statistical Review}, 90(1):118--145.

\bibitem[Gates et~al., 2020]{gates2020latent}
Gates, K.~M., Fisher, Z.~F., and Bollen, K.~A. (2020).
\newblock Latent variable gimme using model implied instrumental variables
  (miivs).
\newblock {\em Psychological methods}, 25(2):227.

\bibitem[Gates et~al., 2017]{gates2017unsupervised}
Gates, K.~M., Lane, S.~T., Varangis, E., Giovanello, K., and Guiskewicz, K.
  (2017).
\newblock Unsupervised classification during time-series model building.
\newblock {\em Multivariate behavioral research}, 52(2):129--148.

\bibitem[Gates and Molenaar, 2012]{gates2012}
Gates, K.~M. and Molenaar, P. C.~M. (2012).
\newblock Group search algorithm recovers effective connectivity maps for
  individuals in homogeneous and heterogeneous samples.
\newblock {\em NeuroImage}, 63(1):310--319.

\bibitem[Hastie et~al., 2015]{hastie2015}
Hastie, T., Tibshirani, R., and Wainwright, M. (2015).
\newblock {\em Statistical learning with sparsity: the lasso and
  generalizations}.
\newblock CRC press.

\bibitem[Henry et~al., 2019]{henry2019}
Henry, T.~R., Feczko, E., Cordova, M., Earl, E., Williams, S., Nigg, J.~T.,
  Fair, D.~A., and Gates, K.~M. (2019).
\newblock Comparing directed functional connectivity between groups with
  confirmatory subgrouping gimme.
\newblock {\em NeuroImage}, 188:642--653.

\bibitem[Hsu et~al., 2008]{hsu2008}
Hsu, N.-J., Hung, H.-L., and Chang, Y.-M. (2008).
\newblock Subset selection for vector autoregressive processes using lasso.
\newblock {\em Computational Statistics \& Data Analysis}, 52(7):3645--3657.

\bibitem[Hunter, 2023]{hunter2022}
Hunter, M.~D. (2023).
\newblock State space mixture modeling: Finding people with similar patterns of
  change.
\newblock {\em Multivariate Behavioral Research}.

\bibitem[Ji et~al., 2018]{ji2018}
Ji, L., Chow, S.-M., Schermerhorn, A.~C., Jacobson, N.~C., and Cummings, E.~M.
  (2018).
\newblock Handling missing data in the modeling of intensive longitudinal data.
\newblock {\em Structural Equation Modeling: A Multidisciplinary Journal},
  25(5):715--736.

\bibitem[Lafit et~al., 2021]{lafit2021}
Lafit, G., Meers, K., and Ceulemans, E. (2021).
\newblock A systematic study into the factors that affect the predictive
  accuracy of multilevel var (1) models.
\newblock {\em Psychometrika}, pages 1--45.

\bibitem[Lane et~al., 2019]{gimme}
Lane, S., Gates, K., Fisher, Z., Arizmendi, C., and Molenaar, P. (2019).
\newblock {\em gimme: Group Iterative Multiple Model Estimation}.
\newblock R package version 0.6-1.

\bibitem[Lane and Gates, 2017]{lane2017}
Lane, S.~T. and Gates, K.~M. (2017).
\newblock Automated selection of robust individual-level structural equation
  models for time series data.
\newblock {\em Structural Equation Modeling: A Multidisciplinary Journal},
  24(5):768--782.

\bibitem[Li et~al., 2022]{li2022}
Li, Y., Wood, J., Ji, L., Chow, S.-M., and Oravecz, Z. (2022).
\newblock Fitting multilevel vector autoregressive models in stan, jags, and
  mplus.
\newblock {\em Structural Equation Modeling: A Multidisciplinary Journal},
  29(3):452--475.

\bibitem[Liu, 2017]{liu2017}
Liu, S. (2017).
\newblock Person-specific versus multilevel autoregressive models: Accuracy in
  parameter estimates at the population and individual levels.
\newblock {\em British Journal of Mathematical and Statistical Psychology},
  70(3):480--498.

\bibitem[Liu, 2018]{liu2018}
Liu, S. (2018).
\newblock Accuracy and reliability of autoregressive parameter estimates: A
  comparison between person-specific and multilevel modeling approaches.
\newblock In {\em Quantitative Psychology: The 82nd Annual Meeting of the
  Psychometric Society, Zurich, Switzerland, 2017}, pages 385--394. Springer.

\bibitem[Loh and Wainwright, 2012a]{loh2012a}
Loh, P.-L. and Wainwright, M.~J. (2012a).
\newblock High-dimensional regression with noisy and missing data: Provable
  guarantees with nonconvexity.
\newblock {\em Annals of Statistics}, 40(3):1637--1664.

\bibitem[Loh and Wainwright, 2012b]{loh2012b}
Loh, P.-L. and Wainwright, M.~J. (2012b).
\newblock Supplement to "{H}igh-dimensional regression with noisy and missing
  data: Provable guarantees with nonconvexity".
\newblock {\em Annals of Statistics}.

\bibitem[Luo et~al., 2022]{luo2022}
Luo, L., Fisher, Z.~F., Arizmendi, C., Molenaar, P., Beltz, A., and Gates,
  K.~M. (2022).
\newblock Estimating both directed and undirected contemporaneous relations in
  time series data using hybrid-group iterative multiple model estimation.
\newblock {\em Psychological Methods}.

\bibitem[L{\"u}tkepohl, 2007]{lutkepohl2007}
L{\"u}tkepohl, H. (2007).
\newblock {\em New {{Introduction}} to {{Multiple Time Series Analysis}}}.
\newblock {Springer Science \& Business Media}.

\bibitem[L{\"u}tkepohl, 2013]{lutkepohl2013}
L{\"u}tkepohl, H. (2013).
\newblock {\em Introduction to multiple time series analysis}.
\newblock Springer Science \& Business Media.

\bibitem[Molenaar et~al., 1993]{molenaar1993}
Molenaar, P., Boomsma, D.~I., and Dolan, C.~V. (1993).
\newblock A third source of developmental differences.
\newblock {\em Behavior genetics}, 23(6):519--524.

\bibitem[Molenaar, 2004]{molenaar2004}
Molenaar, P.~C. (2004).
\newblock A manifesto on psychology as idiographic science: Bringing the person
  back into scientific psychology, this time forever.
\newblock {\em Measurement}, 2(4):201--218.

\bibitem[Nesselroade, 2007]{nesselroade2007}
Nesselroade, J.~R. (2007).
\newblock Factoring at the {Individual} {Level}: {Some} {Matters} for the
  {Second} {Century} of {Factor} {Analysis}.
\newblock In Cudeck, R. and MacCallum, R.~C., editors, {\em Factor {Analysis}
  at 100: {Historical} {Developments} and {Future} {Directions}}. Routledge,
  Mahwah, N.J., 1 edition edition.

\bibitem[Nunes et~al., 2020]{nunes2020}
Nunes, A., Trappenberg, T., and Alda, M. (2020).
\newblock The definition and measurement of heterogeneity.
\newblock {\em Translational psychiatry}, 10(1):299.

\bibitem[Ollier and Viallon, 2017]{ollier2017}
Ollier, E. and Viallon, V. (2017).
\newblock Regression modelling on stratified data with the lasso.
\newblock {\em Biometrika}, 104(1):83--96.

\bibitem[Park et~al., 2022a]{park2022b}
Park, J., Chow, S.-M., Epskamp, S., and Molenaar, P. (2022a).
\newblock Subgrouping with chain graphical var models.

\bibitem[Park et~al., 2022b]{park2022a}
Park, J., Fisher, Z., Chow, S.-M., and Molenaar, P. (2022b).
\newblock On subgrouping continuous processes in discrete time.

\bibitem[Power et~al., 2011]{power2011}
Power, J.~D., Cohen, A.~L., Nelson, S.~M., Wig, G.~S., Barnes, K.~A., Church,
  J.~A., Vogel, A.~C., Laumann, T.~O., Miezin, F.~M., Schlaggar, B.~L., et~al.
  (2011).
\newblock Functional network organization of the human brain.
\newblock {\em Neuron}, 72(4):665--678.

\bibitem[Sims, 1980]{sims1980}
Sims, C.~A. (1980).
\newblock Macroeconomics and {{Reality}}.
\newblock {\em Econometrica}, 48(1):1--48.

\bibitem[Song and Bickel, 2011]{song2011}
Song, S. and Bickel, P.~J. (2011).
\newblock Large {{Vector Auto Regressions}}.
\newblock {\em arXiv:1106.3915 [q-fin, stat]}.

\bibitem[S{\"o}rbom, 1989]{sorbom1989}
S{\"o}rbom, D. (1989).
\newblock Model modification.
\newblock {\em Psychometrika}, 54(3):371--384.

\bibitem[Stern et~al., 2012]{stern2012resting}
Stern, E.~R., Fitzgerald, K.~D., Welsh, R.~C., Abelson, J.~L., and Taylor,
  S.~F. (2012).
\newblock Resting-state functional connectivity between fronto-parietal and
  default mode networks in obsessive-compulsive disorder.
\newblock {\em PloS one}, 7(5):e36356.

\bibitem[Takano et~al., 2021]{takano2021}
Takano, K., Stefanovic, M., Rosenkranz, T., and Ehring, T. (2021).
\newblock Clustering individuals on limited features of a vector autoregressive
  model.
\newblock {\em Multivariate Behavioral Research}, 56(5):768--786.

\bibitem[Tibshirani, 1996]{tibshirani1996}
Tibshirani, R. (1996).
\newblock Regression {{Shrinkage}} and {{Selection}} via the {{Lasso}}.
\newblock {\em Journal of the Royal Statistical Society. Series B
  (Methodological)}, 58(1):267--288.

\bibitem[Tricomi et~al., 2009]{tricomi2009}
Tricomi, E., Balleine, B.~W., and O’Doherty, J.~P. (2009).
\newblock A specific role for posterior dorsolateral striatum in human habit
  learning.
\newblock {\em European Journal of Neuroscience}, 29(11):2225--2232.

\bibitem[Uddin et~al., 2009]{uddin2009}
Uddin, L.~Q., Clare~Kelly, A., Biswal, B.~B., Xavier~Castellanos, F., and
  Milham, M.~P. (2009).
\newblock Functional connectivity of default mode network components:
  correlation, anticorrelation, and causality.
\newblock {\em Human brain mapping}, 30(2):625--637.

\bibitem[Wang et~al., 2022]{wang2022changes}
Wang, X., Zwosta, K., Wolfensteller, U., and Ruge, H. (2022).
\newblock Changes in global functional network properties predict individual
  differences in habit formation.
\newblock {\em Human Brain Mapping}.

\bibitem[Zellner, 1962]{zellner1962}
Zellner, A. (1962).
\newblock An {{Efficient Method}} of {{Estimating Seemingly Unrelated
  Regressions}} and {{Tests}} for {{Aggregation Bias}}.
\newblock {\em Journal of the American Statistical Association},
  57(298):348--368.

\bibitem[Zhao and Yu, 2006]{zhao2006}
Zhao, P. and Yu, B. (2006).
\newblock On model selection consistency of lasso.
\newblock {\em The Journal of Machine Learning Research}, 7:2541--2563.

\bibitem[Zhou et~al., 2009]{zhou2009}
Zhou, S., van~de Geer, S., and B{\"u}hlmann, P. (2009).
\newblock Adaptive lasso for high dimensional regression and gaussian graphical
  modeling.
\newblock {\em arXiv preprint arXiv:0903.2515}.

\bibitem[Zilverstand and Goldstein, 2020]{zilverstand2020dual}
Zilverstand, A. and Goldstein, R.~Z. (2020).
\newblock Dual models of drug addiction: The impaired response inhibition and
  salience attribution model.
\newblock In {\em Cognition and addiction}, pages 17--23. Elsevier.

\bibitem[Zou, 2006]{zou2006}
Zou, H. (2006).
\newblock The {{Adaptive Lasso}} and {{Its Oracle Properties}}.
\newblock {\em Journal of the American Statistical Association},
  101(476):1418--1429.

\end{thebibliography}
\end{singlespace}
\end{document}